\documentclass[12pt]{article}
\usepackage{amsmath,amssymb,cite}
\catcode`\@=11

\@addtoreset{equation}{section}
\def\theequation{\arabic{section}.\arabic{equation}}
     
\catcode`\@=11
\def\thesection{\arabic{section}}

\def\appendix{\setcounter{section}{0}
        \def\thesection{Appendix.}
        \def\theequation{\Alph{section}.\arabic{equation}}}
\def\section{\@startsection{section}{1}{\z@}{3.5ex plus 1ex minus
   .2ex}{2.3ex plus .2ex}{\large\bf}}

\newcommand{\captionfonts}{\small}

\makeatletter  
\long\def\@makecaption#1#2{%
  \vskip\abovecaptionskip
  \sbox\@tempboxa{{\captionfonts #1: #2}}%
  \ifdim \wd\@tempboxa >\hsize
    {\captionfonts #1: #2\par}
  \else
    \hbox to\hsize{\hfil\box\@tempboxa\hfil}%
  \fi
  \vskip\belowcaptionskip}
\makeatother   
\newcommand{\Pre}{{\it Preprint\ }}

\long\def\@makefntext#1{\parindent 0cm\noindent
\hbox to 1em{\hss$^{\@thefnmark}$}#1}

\begin{document}
\begin{titlepage}
\vspace{.5in}
\begin{flushright}
gr-qc/0601041\\
December 2005\\
\end{flushright}
\vspace{.5in}
\begin{center}
{\Large\bf
  Horizons, Constraints, and Black Hole Entropy}\\
\vspace{.4in}
{S.~C{\sc arlip}\footnote{\it email: carlip@physics.ucdavis.edu}\\
       {\small\it Department of Physics}\\
       {\small\it University of California}\\
       {\small\it Davis, CA 95616}\\{\small\it USA}}
\end{center}

\vspace{.5in}
\begin{center}
{\large\bf Abstract}
\end{center}
\begin{center}
\begin{minipage}{5.25in}
{\small
Black hole entropy appears to be ``universal''---many independent  
calculations, involving models with very different microscopic
degrees of freedom, all yield the same density of states.  I discuss
the proposal that this universality comes from the behavior of  
the underlying symmetries of the classical theory.  To impose the 
condition that a black hole be present, we must partially break 
the classical symmetries of general relativity, and the resulting
Goldstone boson-like degrees of freedom may account for the
Bekenstein-Hawking entropy.  In particular, I demonstrate that 
the imposition of a ``stretched horizon'' constraint modifies the 
algebra of symmetries at the horizon, allowing the use of standard
conformal field theory techniques to determine the asymptotic 
density of states.  The results reproduce the Bekenstein-Hawking 
entropy without any need for detailed assumptions about the 
microscopic theory. 
}
\end{minipage}
\end{center}
\end{titlepage}
\addtocounter{footnote}{-1}

\section{Introduction}

In the continuing quest for a quantum theory of gravity, black hole 
thermodynamics may be the nearest thing we have to ``experimental''
data.  Hawking radiation has not been directly observed, of
course.  But the thermal properties of black holes have been 
derived in so many independent ways \cite{Hawking,GibHawk,%
GibPerry,Haag,KW,Unruh,DeWitt,Parikh,Nadalini}, and are so robust 
against changes in the starting assumptions \cite{Barcelo,Unruh2}, 
that it would seem perverse to devote too much time to a putative 
quantum theory of gravity that could not reproduce the standard 
results.  

Since the Bekenstein-Hawking entropy 
\begin{equation}
S = \frac{A}{4\hbar G}
\label{a0}
\end{equation}
depends on both Planck's constant and Newton's constant, it is
inherently quantum gravitational.  We might therefore hope that 
black hole thermodynamics could give us important clues to the 
question of how to quantize general relativity.  The fundamental 
problem is to understand the underlying statistical mechanics.
What microscopic states are responsible for the Hawking temperature
and Bekenstein-Hawking entropy?  

Ten years ago, this question has an almost universally accepted
answer: we don't know.  There were some interesting ideas floating
around, involving entanglement entropy \cite{Sorkin} and entropy 
of an ``atmosphere'' of external fields near the horizon \cite{tHooft}, 
but we had nothing close to a complete description.

Today, things are radically different: many people will tell you 
with great confidence exactly what microscopic physics leads to 
black hole thermodynamics.  The new problem is that while many people 
``know'' the answer, they do not all agree.  Black hole entropy may come 
from weakly coupled string and D-brane states \cite{StromVafa,Peet}; 
from nonsingular string ``fuzzballs'' \cite{Mathur}; from holographic
states in a dual conformal field theory that is in some sense located
at infinity \cite{AGMOO}; from spin network states at the horizon 
\cite{Ashtekar} or perhaps inside the horizon \cite{X}; from ``heavy'' 
degrees of freedom in induced gravity \cite{Fursaev}; or perhaps from 
nonlocal topological properties of the black hole spacetime \cite{Hawkingb}.  
So far none of these pictures offers us a comprehensive picture of black 
hole thermodynamics.  But in its realm of applicability, each can be 
used to count states for some black holes, and each seems to give the 
correct entropy (\ref{a0}).  

In a field in which we do not yet know the answers, the existence of 
competing models may be seen as a sign of health.  But the existence 
of competing models that all \emph{agree} cries out for a deeper 
explanation.    

\section{Conformal symmetry and state-counting \label{sae}}

While there may be others, I know of only one general approach that might 
explain this universality: perhaps some underlying feature of classical 
general relativity constrains the structure of any good quantum theory 
of gravity.  The natural candidate for such a feature is a symmetry.  
This may at first seem unlikely---entropy is determined by the density 
of states, an inherently quantum mechanical characteristic, and it is 
not at all obvious that a classical symmetry can provide a strong enough 
constraint on the quantum theory.  In one case, though, a known classical 
symmetry does just what we need.

Start with a two-dimensional conformal field theory, that is, a theory
invariant under diffeomorphisms and Weyl transformations, and choose  
complex coordinates $z$ and ${\bar z}$.  The fundamental symmetries
of such a theory are holomorphic and antiholomorphic diffeomorphisms, 
which are canonically generated by ``Virasoro generators'' $L[\xi]$ 
and ${\bar L}[{\bar \xi}]$ \cite{CFT}.  Such a theory has two conserved 
charges, $L_0 = L[\xi_0]$ and ${\bar L}_0 = {\bar L}[{\bar\xi}_0]$, 
which can be thought of as ``energies'' with respect to constant 
holomorphic and antiholomorphic transformations, or as equivalently
as linear combinations of energy and angular momentum.

As generators of diffeomorphisms, the Virasoro generators have an algebra 
that is almost unique \cite{Teitelboim}: 
\begin{align}
&\left\{L[\xi],L[\eta]\right\} = L[\eta\xi' - \xi\eta']
  + \frac{c}{48\pi}\int dz\left( \eta'\xi'' - \xi'\eta''\right) \nonumber \\
&\left\{L[\xi],{\bar L}[{\bar\eta}]\right\} = 0 \label{a1} \\
&\left\{{\bar L}[{\bar\xi}],{\bar L}[{\bar\eta}]\right\} 
  = {\bar L}[{\bar\eta}{\bar\xi'} - {\bar\xi}{\bar\eta'}]
  + \frac{{\bar c}}{48\pi}\int d{\bar z}\left( {\bar\eta}'{\bar\xi}'' 
  - {\bar\xi}'{\bar\eta}''\right)\nonumber .
\end{align}
The central charges (also known as ``conformal anomalies'') $c$ and 
$\bar c$ determine the unique central extension of the ordinary algebra 
of diffeomorphisms.  These constants can occur classically, coming, 
for instance, from boundary terms \cite{BH}, and even if they are 
absent in the classical theory, they will typically appear upon 
quantization.

Consider now a conformal field theory for which the lowest eigenvalues
of $L_0$ and ${\bar L}_0$ are nonnegative numbers $\Delta_0$ and 
${\bar\Delta}_0$.  Cardy has shown \cite{Cardy,Cardyb} that for large 
eigenvalues $\Delta$ and $\bar\Delta$ of $L_0$ and ${\bar L}_0$, the 
density of states $\rho(\Delta,\bar\Delta)$ takes the remarkably simple 
form
\begin{equation}
\ln\rho(\Delta,\bar\Delta) \sim 2\pi\left\{
  \sqrt{\frac{c_{\hbox{\scriptsize\it eff}}\Delta}{6}} 
  + \sqrt{\frac{{\bar c}_{\hbox{\scriptsize\it eff}}{\bar\Delta}}{6}}\,\right\}, 
  \quad \hbox{with}\ \ 
  c_{\hbox{\scriptsize\it eff}} = c-24\Delta_0, \ 
  {\bar c}_{\hbox{\scriptsize\it eff}} = {\bar c}-24{\bar\Delta}_0 .
\label{a2}
\end{equation}
The entropy is thus determined by the symmetry, independent of any other 
details.  In particular, two conformal field theories with very different
field content will have the same asymptotic density of states, provided that
their effective central charges are equal.

The Cardy formula (\ref{a2}) is relatively straightforward to prove---see,
for example, \cite{Carlipy}---but I do 
not know of a fundamental physical explanation for this result.  Partial 
insight may be obtained, though, by noting that a central charge reflects a  
breaking of the conformal symmetry.  As we know from field theory \cite{Goldstone}, 
a broken symmetry can lead to new ``Goldstone'' degrees of freedom.\footnote{The 
analogy with the Goldstone mechanism was suggested to me by Kaloper and Terning 
\cite{Kaloper}.}  Here, for example, one would normally require that physical 
states be annihilated by the generators $L[\xi]$ and ${\bar L}[{\bar\xi}]$ of 
gauge symmetries,
\begin{equation}
L[\xi]|\hbox{\it phys}\rangle  = {\bar L}[{\bar\xi}]|\hbox{\it phys}\rangle = 0.
\label{a3}
\end{equation}
But if $c\ne0$, these conditions are incompatible with the brackets (\ref{a1}),
and must be relaxed, for instance by requiring that only the positive frequency
components of $L[\xi]$ annihilate physical states.  We thus obtain new physical 
states---``would-be gauge states'' \cite{Carlipe}  that become physical because 
of the broken symmetry---that may contribute to $\rho(\Delta,\bar\Delta)$.

\section{Near-horizon is near-conformal \label{nhnc}}

General relativity is not a conformal field theory, and it is most certainly
not two-dimensional.  It is therefore not obvious that the the Cardy formula 
should have any particular relevance to black hole entropy.  But there are good
reasons to believe that black hole dynamics may be \emph{effectively} described 
by a two-dimensional conformal field theory near the horizon.  

For instance, it is known that near a horizon, matter can be described by a 
two-dimensional conformal field theory, with fields depending only on $t$ and 
the ``tortoise coordinate'' $r_*$ \cite{Birmingham,Gupta,Camblong}.  Indeed, 
the near-horizon metric in such coordinates becomes, in any dimension,
\begin{equation}
ds^2 = N^2(dt^2-dr_*{}^2) + ds_\perp{}^2 ,
\label{ab1}
\end{equation}
where the lapse function $N$ goes to zero at the horizon.  The Klein-Gordon
equation then reduces to
\begin{equation}
(\Box - m^2)\varphi =
     \frac{1}{N^2}(\partial_t^2 - \partial_{r_*}^2)\varphi + \mathcal{O}(1) = 0 .
\label{ab2}
\end{equation}
The mass and transverse excitations become negligible near the horizon: they
are essentially red-shifted away relative to excitations in the $r_*$-$t$
plane, leaving an effective two-dimensional conformal field theory at each 
point of the horizon.  A similar dimensional reduction occurs for the Dirac
equation. 

Jacobson and Kang have observed that the surface gravity and temperature 
of a stationary black hole are conformally invariant as well \cite{Jacobson}.  
And perhaps of most interest for the issues at hand, Medved et al.\ have shown 
that a generic stationary black hole metric always has an approximate conformal 
Killing vector near the horizon \cite{Martin,Martinb}.

\section{The BTZ black hole \label{btz}}

The first solid evidence that the Cardy formula might be used to determine 
black hole thermodynamics came from the (2+1)-dimensional black hole of 
Ba{\~n}ados, Teitelboim, and Zanelli \cite{BTZ,BHTZ,Carlipa,Carlipb}.  The
BTZ black hole is the (2+1)-dimensional analog of the Kerr-AdS geometry,
with a metric of the form 
\begin{align}
ds^2 = N^2dt^2 - &N^{-2}dr^2 - r^2\left( d\phi + N^\phi dt\right)^2
 \nonumber\\
 &\hbox{with}\ \ N 
 = \left( -8GM + \frac{r^2}{\ell^2} + \frac{16G^2J^2}{r^2} \right)^{1/2},\ \
N^\phi = - \frac{4GJ}{r^2} ,\label{b1} 
\end{align}
where the cosmological constant is $\Lambda=-1/\ell^2$ and $M$ and $J$ are the 
anti-de Sitter analogs of ADM mass and angular momentum.  Like all vacuum 
spacetimes in 2+1 dimensions \cite{Carlipc}, the BTZ metric has constant 
curvature, and can  in fact be expressed as a quotient of anti-de Sitter 
space by a discrete group of isometries.  Nevertheless, it is a genuine black 
hole:
\begin{itemize}
\item It has an event horizon at $r=r_+$ and an inner Cauchy
horizon at $r=r_-$, where
\begin{equation}
r_\pm^2=4GM\ell^2\left \{ 1 \pm
\left [ 1 - \left(\frac{J}{M\ell}\right )^2\right ]^{1/2}\right \} ;
\label{b2}
\end{equation}
\item its Carter-Penrose diagram is identical to that of an ordinary Kerr-AdS 
 black hole;  
\item it occurs as the end point of gravitational collapse of matter;
\item most important for us, it exhibits standard black hole 
 thermodynamics, with an entropy (\ref{a0}), where the two-dimensional
 ``area'' is the horizon circumference.
\end{itemize}
 
Despite its apparent simplicity, though, the thermodynamics of the BTZ black
hole presents us with a conundrum.  General relativity in 2+1 dimensions
has no local degrees of freedom \cite{Carlipc}: up to a finite number of global 
degrees of freedom, the metric is completely determined by the constraints.  
There seems to be no room for enough states to account for what can be an 
arbitrarily large entropy.

One piece of the answer was discovered independently in 1997 by Strominger 
\cite{Strominger} and Birmingham, Sachs, and Sen \cite{BSS}.  Note first
that the conformal boundary of any (2+1)-dimensional asymptotically anti-de 
Sitter spacetime is a flat cylinder.  It is thus unsurprising that the asymptotic 
symmetries of the BTZ black hole are described by a Virasoro algebra (\ref{a1}).  
It is a bit more surprising that this algebra has a central extension.  But as 
Brown and Henneaux showed in 1986 \cite{BH}---and many authors have subsequently 
confirmed \cite{Carlipb}---a classical central charge is in fact present,
appearing because of the need to add boundary terms to the constraints, and
taking the value
\begin{equation}
c = \frac{3\ell}{2\hbar G} .
\label{b3}
\end{equation}
Furthermore, the two conserved charges $L_0$ and ${\bar L}_0$ can be computed by
standard ADM methods, yielding
\begin{equation}
\Delta_0\sim\frac{1}{16\hbar G\ell} (r_+ + r_-)^2, \quad 
{\bar\Delta}_0\sim\frac{1}{16\hbar G\ell} (r_+ - r_-)^2 .
\label{b4}
\end{equation}
Inserting (\ref{b3})--(\ref{b4}) into the Cardy formula, we find
\begin{equation}
S \sim \frac{2\pi}{8\hbar G}(r_+ + r_-) + \frac{2\pi}{8\hbar G}(r_+ - r_-) 
  = \frac{2\pi r_+}{4\hbar G} ,
\label{b5}
\end{equation}
the correct Bekenstein-Hawking entropy.  We thus learn that the BTZ entropy is 
related to symmetries and boundary conditions at infinity.

A second piece of the answer starts from the observation \cite{Achucarro,Witten}
that (2+1)-dimensional gravity with a negative cosmological constant can be 
written as an $\mathrm{SO}(2,1)\times\mathrm{SO}(2,1)$ Chern-Simons theory.  The 
Chern-Simons action is gauge invariant on a compact manifold.  But boundaries 
and boundary conditions can break that invariance, in much the way I discussed 
in section \ref{sae}.  The resulting Goldstone-like modes are described by a 
two-dimensional conformal field theory at infinity: a Wess-Zumino-Witten model 
\cite{Wittenb,EMSS,Carlipm}, or, in the case of (2+1)-dimensional gravity, a 
Liouville theory with central charge (\ref{b3}) \cite{CHvD}.  

This result has now been confirmed by many different approaches, ranging from 
Chern-Simons methods to the AdS/CFT correspondence; see \cite{Carlipb} for a
review.  Of particular interest is the explicit derivation of the Liouville
field as a ``would-be diffeomorphism'' that becomes dynamical because of the 
need to impose boundary conditions \cite{Manvelyan,Carlipd}.  This appears to
resolves our earlier paradox: although (2+1)-dimensional gravity on a compact 
manifold has only a small number of topological degrees of freedom, the presence 
of a boundary partially breaks the diffeomorphism invariance, promoting ``gauge'' 
degrees of freedom to physical excitations.
  
Whether these Liouville degrees of freedom can account for the entropy 
(\ref{b5}) remains an open question \cite{Carlipb}.  A naive application 
of the Cardy formula certainly yields the correct entropy.  The relatively 
well-understood ``normalizable'' sector of Liouville theory has a nonzero 
minimum eigenvalue $\Delta_0$ of $L_0$, however, lowering the effective 
central charge $c_{\hbox{\scriptsize\it eff}}$ in (\ref{a1}) and ruining the 
correspondence \cite{Kutasov,Martinec}.  Chen has recently shown that the more 
poorly understood ``nonnormalizable'' sector of the theory may admit a 
sensible quantization, though, with states that can be explicitly 
counted and that seem to reproduce the correct entropy \cite{Chen}.

\section{Horizons and constraints \label{hc}}

The real world, of course, is not 2+1 dimensional, and the derivation of the 
BTZ black hole entropy does not easily generalize.  Only in 2+1 dimensions 
is the asymptotic boundary of spacetime two dimensional, allowing a direct 
application of the Cardy formula.  Moreover, it seems more natural to associate 
black hole states with the horizon rather than spatial infinity.  For a 
single black hole in 2+1 dimensions, the distinction may be unimportant, 
since there are no additional dynamical degrees of freedom lying between 
the horizon and infinity.  In 3+1 dimensions, though, the prospect of 
extracting the degrees of freedom of a single black hole from the asymptotics 
of a complicated spacetime is daunting.

We can, however, draw a few lessons from 2+1 dimensions.  We should look for 
``broken gauge invariance'' from boundary conditions, and hope for an 
effective two-dimensional picture.  But we should also start by looking near 
the horizon rather than at infinity.

There is an immediate objection to this proposal: unlike spatial infinity, 
the horizon of a black hole is not a true boundary.  To clarify this issue, 
we must take a step back and ask what it means to ask a question about a black 
hole in a quantum theory of gravity.

In the usual semiclassical treatments of black hole thermodynamics, the answer 
to this question is obvious: we simply fix a black hole background, and look 
at quantum fields in that background.  In a full quantum theory of gravity, 
though, we are not allowed to do this.  Such a theory has no fixed background; 
the metric is an operator, and the uncertainty principle tells us that its value 
cannot be exactly specified. At best, we can restrict a piece of the geometry 
and ask \emph{conditional} questions: ``If geometric features that characterize 
a black hole of type X are present, what is the probability of observing 
phenomenon Y?''

I know of two ways to obtain such a conditional probability.  The first, 
discussed in \cite{Carlipe}, is to treat the horizon as a ``boundary'' at 
which suitable boundary conditions are imposed.  In the sum over histories 
formalism, for example, we can divide spacetime into two regions along a 
hypersurface $\mathcal H$ and perform separate path integrals over fields 
on each side, with fields restricted at the ``boundary'' by the requirement 
that $\mathcal H$ be a horizon.  Such split path integral has been studied 
in detail in 2+1 dimensions \cite{Wittenc}, where it leads to the same WZW 
model that was discussed in the preceding section.  Although the horizon is 
not a true boundary, it is, in this approach, a hypersurface upon which 
we impose ``boundary conditions,'' and this turns out to be good enough.

Alternatively, we can impose ``horizon constraints'' directly, either classically
or in the quantum theory.  We might, for example, construct an operator $\vartheta$
representing the expansion of a particular null surface, and restrict ourselves 
to states annihilated by $\vartheta$.  As we shall see below, such a restriction 
can affect the algebra of diffeomorphisms, allowing us to exploit the Cardy formula
to count states.

The ``horizon as a boundary'' approach has been widely investigated; see, for 
instance, \cite{Carliph,Carlipi,Navarro,Jing,Izq,Park,Silva,Cvitan,Cvitanb,Cadoni}.  
One naturally finds a conformal symmetry in the $r_*$--$t$ plane, and one can 
obtain a Virasoro algebra with a central charge that leads to the correct black hole 
entropy.  On the other hand, the diffeomorphisms whose algebra yields that central 
charge---essentially those that leave the lapse function invariant---are generated 
by vector fields that blow up at the horizon \cite{Dreyer,Koga,Pinamonti,Huang}, 
and it is not clear whether this is permissible.  Nor is the angular dependence
of these vector fields well understood.  A related approach looks for approximate 
conformal symmetry near the horizon \cite{Solo,Carlipj,Giacomini,Solob}; again, 
one finds a Virasoro algebra with a central charge that seems to lead to the 
correct entropy.

The alternative ``horizon constraint'' approach is still in the early stages of 
development.  Suppose we wish to constrain our theory of gravity by requiring 
that some prescribed surface $\mathcal H$ be an ``isolated horizon'' \cite{Ashtekarb} 
or a ``dynamical horizon'' \cite{Booth}.  Such constraints restrict the allowed 
data on $\mathcal H$, and in principle we should be able to use the tools of 
constrained Hamiltonian dynamics \cite{Dirac,Diracb,Bergmann} to study such 
conditions.  Unfortunately, though, an isolated horizon is by definition a 
null surface, while standard techniques deal with constraints on spacelike
surfaces.  While some work on ``constrained light cone quantization''  exists,
this is a difficult program.

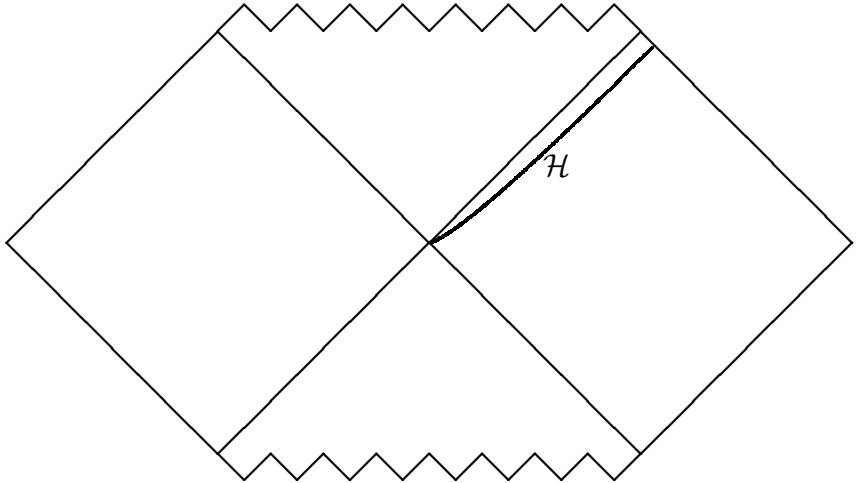
\begin{figure}
\centering
\begin{picture}(200,160)(32,-5)
\thicklines
%
%
\put(50,0){\line(1,1){160}}
\put(50,160){\line(1,-1){160}}
\put(50,0){\line(-1,1){80}}
\put(-30,80){\line(1,1){80}}
\put(210,0){\line(1,1){80}}
\put(290,80){\line(-1,1){80}}
\multiput(50,160)(20,0){8}{\line(1,1){10}}
\multiput(60,170)(20,0){8}{\line(1,-1){10}}
\multiput(50,0)(20,0){8}{\line(1,-1){10}}
\multiput(60,-10)(20,0){8}{\line(1,1){10}}
\qbezier(130,80)(145,85)(214,154)
\put(173,105){$\mathcal H$}
\end{picture}
\caption{A spacelike ``stretched horizon'' $\mathcal H$ \label{fig1}}
\end{figure}

As a simpler warm-up exercise, we can impose constraints requiring the presence 
of a spacelike ``stretched horizon'' that becomes nearly null, as illustrated 
in figure \ref{fig1}.  As I will discuss below, such a stretched horizon constraint 
leads to a Virasoro algebra with a calculable central charge, and at least in 
the case of two-dimensional dilaton gravity, yields the correct Bekenstein-Hawking 
entropy.

\section{Horizon constraints and the dilaton black hole \label{hcd}}

I now turn to a slightly more technical analysis of a particular case, 
the two-dimensional dilaton black hole.  Details can be found in 
\cite{Carlipf,Carlipg}.  Our world is surely not two dimensional, so 
this may  again seem too great a specialization.  As argued above, 
though, we expect the near-horizon dynamics of an arbitrary black hole 
to be effectively two dimensional, so a dimensionally reduced model is 
not such a bad starting point.

Two-dimensional dilaton gravity \cite{Kunstatter,Grumiller} can be described,
after a rescaling of the metric, by an action 
\begin{equation}
I = \int d^2x \sqrt{-g}\left[AR + V(A)\right] ,
\label{c1}
\end{equation}
where $R$ is the two-dimensional scalar curvature and $A$ is a scalar field, 
the dilaton (often denoted as $\varphi$).  $V(A)$ is a potential whose 
form depends on the higher-dimensional theory we started with; we will not 
need an exact expression. As the notation suggests, $A$ is the transverse 
area in the higher-dimensional theory, in units $16\pi G=1$.  The 
analog of the expansion---the fractional rate of change of area along a 
null curve with null normal $l^a$---is
\begin{equation}
\vartheta = l^a\nabla_aA/A .
\label{c1a}
\end{equation}

It is useful to rewrite the action (\ref{c1}) in terms of a null dyad $(l^a,n^a)$
with $l^2=n^2=0$, $l\cdot n=-1$.  These determine ``surface gravities'' $\kappa$ 
and $\bar\kappa$, defined by the conditions
\begin{equation}
\nabla_al_b = -\kappa n_al_b - {\bar\kappa}l_al_b , \qquad
\nabla_an_b = \kappa n_an_b + {\bar\kappa}l_an_b ,
\label{c2}
\end{equation}
and the action becomes
\begin{equation}
I =  \int d^2x \left[{\hat\epsilon}^{ab}
     \left(2\kappa n_b\partial_aA
     - 2{\bar\kappa} l_b\partial_aA\right) + \sqrt{-g}V\right] .
\label{c3}
\end{equation}
If we now express the components of our dyad with respect to coordinates $(u,v)$ as
\begin{equation}
l = \sigma du + \alpha dv, \qquad n = \beta du + \tau dv ,
\label{c4}
\end{equation}
it is easy to find the Hamiltonian form of the action \cite{Carlipf}.  The 
system has three first-class constraints; denoting a derivative with respect 
to $v$ by a prime, they are
\begin{eqnarray} 
&&C_\perp = \pi_\alpha{}' - \frac{1}{2}\pi_\alpha\pi_A - \tau V(A) 
 \nonumber\\
&&C_\parallel = \pi_A A' - \alpha\pi_\alpha{}' - \tau\pi_\tau{}' \vphantom{\frac{1}{2}}
 \label{c5}\\
&&C_\pi = \tau\pi_\tau - \alpha\pi_\alpha + 2A' \vphantom{\frac{1}{2}} .\nonumber
\end{eqnarray}
$C_\perp$ and $C_\parallel$ generate the diffeomorphisms orthogonal to and parallel 
to a spacelike slice $u=\mathit{const}.$, while $C_\pi$ generates local Lorentz 
transformations.

We can now impose ``stretched horizon'' constraints at the surface $u=0$.  
We first demand that $\mathcal H$ be ``almost null,'' i.e., that its normal be 
nearly equal to the null vector $l^a$.  By (\ref{c4}), this requires that 
$\alpha=\epsilon_1\ll1$.  

We must next demand that $\mathcal H$ be ``almost nonexpanding.''  This is a bit 
more subtle, since the absolute scale of $l^a$ is not fixed; while the requirement 
that the expansion be exactly zero is independent of such a scale, the requirement 
of a ``small'' expansion is not.  One solution is to note that the surface gravity 
$\kappa$ depends on this scale as well, and that the ratio $\vartheta/\kappa A$ is 
independent of at least constant rescalings of $l^a$.  We therefore require that 
$l^v\nabla_vA/\kappa A= \epsilon_2\ll1$.  Expressing these conditions in terms of 
canonical variables, we obtain two ``stretched horizon'' constraints:
\begin{eqnarray}
&&K_1 = \alpha-\epsilon_1 = 0 \nonumber\\
&&K_2 = A' - \frac{1}{2}\epsilon_2A_+\pi_A + \frac{a}{2}C_\pi = 0 ,
\label{c6}
\end{eqnarray}
where $a$ is an arbitrary constant and $A_+$ is the horizon value of the dilaton.
By looking at a generic exact black hole solution, one can verify that these 
constraints do, in fact, determine a spacelike stretched horizon that looks like 
that of figure \ref{fig1}, which very rapidly becomes very nearly null.

$K_1$ and $K_2$ are not quite ``constraints'' in the usual sense of constrained
Hamiltonian dynamics, but they are similar enough that many existing techniques 
can be used.  In particular, observe that the $K_i$ have nontrivial brackets with 
the momentum and boost generators $C_\parallel$ and $C_\pi$, so these no longer 
generate invariances of the constrained theory. But we can fix this by a method
suggested years ago by Bergmann and Komar \cite{Bergmann}: we define new generators
\begin{eqnarray}
&&C_\parallel \rightarrow C_\parallel^* = C_\parallel + a_1K_1 + a_2K_2 \nonumber\\
&&C_\pi \rightarrow C_\pi^* = C_\pi + b_1K_1 + b_2K_2 
\label{c7}
\end{eqnarray}
with coefficients $a_i$ and $b_i$ chosen so that $\{C^*,K_i\}=0$.  Since $K_i=0$
on admissible geometries, the generators $C^*$ are physically equivalent to the 
original $C$; but they now preserve the horizon constraints as well.   

We now make the crucial observation that the redefinitions (\ref{c7}) affect the 
Poisson brackets of the constraints.  With the choice $a=-2$ in (\ref{c6}), it
may be shown that
\begin{eqnarray}
&&\{C_\parallel^*[\xi],C_\parallel^*[\eta]\} = -C_\parallel^*[\xi\eta'-\eta\xi'] 
  + \frac{1}{2}\epsilon_2A_+\int\,dv(\xi'\eta'' - \eta'\xi'') \nonumber\\
&&\{C_\parallel^*[\xi],C_\pi^*[\eta]\} = -C_\pi^*[\xi\eta'] \label{c8}\\
&&\{C_\pi^*[\xi],C_\pi^*[\eta]\} = -\frac{1}{2}\epsilon_2A_+\int\,dv(\xi\eta' - \eta\xi') .
  \nonumber
\end{eqnarray}
This algebra has a simple conformal field theoretical interpretation 
\cite{CFT}: the $C_\parallel^*$ generate a Virasoro algebra with central charge
\begin{equation}
\frac{c}{48\pi} = -\frac{1}{2}\epsilon_2A_+ ,
\label{c9}
\end{equation}
while $C_\pi^*$ is a primary field of weight one.  Different choices of the parameter
$a$ in (\ref{c6}) yield equivalent algebras, although with slightly redefined 
generators.

The Cardy formula (\ref{a2}) requires both the central charge and the conserved 
charge $\Delta$.  As in the usual approaches to black hole mechanics, the latter 
comes from a boundary term needed to make the generator $C_\parallel^*$ 
``differentiable'' \cite{Regge}.  It was shown in \cite{Carlipf} that this term is
\begin{equation}
C_{\parallel\,\mathit{bdry}}^*[\xi] = -\left.\xi\pi_AA\right|_{v=v_+} ,
\label{c10}
\end{equation}
which will give a nonvanishing classical contribution to $\Delta$.

Finally, we also need a mode expansion to define the Fourier component 
$L_0$, or, equivalently, a normalization for the ``constant translation'' 
$\xi_0$.  For a conformal field theory defined on a circle, or on a full 
complex plane with a natural complex coordinate, this normalization is 
essentially unique.  Here, though, it is not so obvious how to choose the 
``right'' complex coordinate.  As argued in \cite{Carlipj}, however, there 
is one particularly natural choice,
\begin{equation}
z = e^{2\pi i A/A_+}, \qquad \xi_n = \frac{A_+}{2\pi A'} z^n,
\label{c11}
\end{equation}
where the prefactor is fixed by demanding that $[\xi_m,\xi_n] = i(n-m)\xi_{m+n}$.

Equation (\ref{c10}) then implies that
\begin{equation}
\Delta = C_{\parallel\,\mathit{bdry}}^*[\xi_0] = -\frac{A_+}{2\pi A'}\pi_AA_+
 = -\frac{A_+}{\pi\epsilon_2} .
\label{c12}
\end{equation}
Inserting (\ref{c9}) and (\ref{c12}) into the Cardy formula, assuming that
$\Delta_0$ is small, and restoring the factors of $16\pi G$ and $\hbar$, we 
obtain an entropy
\begin{equation}
S = \frac{2\pi}{16\pi G}\sqrt{\left(-\frac{24\pi\epsilon_2A_+}{6\hbar}\right)
  \left(-\frac{A_+}{\pi\epsilon_2\hbar}\right)} = \frac{A_+}{4\hbar G} ,
\label{c13}
\end{equation}
exactly reproducing the standard Bekenstein-Hawking entropy (\ref{a0}).

\section{Open questions}

While these results are intriguing, they are certainly not yet conclusive.  
I know of several straightforward steps that may take us further:
\begin{enumerate} 
\item We should determine how sensitive the result is to the exact definition
(\ref{c6}) of the stretched horizon $\mathcal H$.  As a first step, a similar 
computation has now been carried out in radially quantized Euclidean quantum 
gravity \cite{Carlipk}.  Here, the constraints have a simpler geometric 
interpretation---they essentially fix the proper distance of the initial
surface from the origin---and it may be possible to relate the constraint
analysis to the path integral methods of \cite{Carlipl}.  Ideally, the 
analysis should also be repeated in a true light cone quantization; work on 
this is in progress.
\item We should extend the analysis beyond two dimensions.  This is probably
not too hard conceptually, though the technical details may be complicated.
\item We should try to relate the horizon constraint approach of section 
\ref{hcd} to the ``horizon as boundary'' methods described in section \ref{hc}.
A comparison with the near-horizon symmetry approaches of \cite{Solo,Carlipj} 
should be fairly simple: the central charge (\ref{c9}) agrees with that of 
\cite{Carlipj}, and can be made to match that of \cite{Solo} by a choice of 
the parameter $q$ in that paper.  The relation to other boundary approaches
may be more subtle, though it is worth noting that the central charge (\ref{c9}) 
agrees with that of \cite{Carliph} with the interesting choice $T=\vartheta$
for the periodicity of the modes in that paper.
\end{enumerate}

Three other questions are more difficult, but perhaps more profound.  First, 
observe that the vector fields (\ref{c11}) blow up at the horizon, where
$A'\rightarrow0$.  A similar phenomenon occur in the ``horizon as boundary''
approach, as noted in \cite{Dreyer,Pinamonti}.  This divergence seems to 
be closely related to the use of Schwarzschild-like coordinates, which 
characterize an observer who stays outside of the black hole.  Note that
such coordinates were needed for the conformal behavior described in section
\ref{nhnc}.  Similarly, an exact horizon constraint has recently been 
analyzed in two-dimensional dilaton gravity, with diffeomorphisms that are 
implicitly required to be well-behaved at the horizon \cite{Grumiller2}; 
for such diffeomorphisms, the central extension of the Virasoro algebra seems 
to disappear.  If this proves to be a general feature of conformal methods, 
it may be telling us something profound about ``black hole complementarity'' 
\cite{Susskind}: perhaps the Bekenstein-Hawking entropy is only well-defined 
for an observer who remains outside the horizon.

Second, if the horizon symmetry described here provides a universal 
explanation of black hole entropy, then the symmetry should be identifiable 
in other approaches to black hole microphysics.  There is a reasonable chance 
that this connection can be made for a large class of ``stringy'' black holes.  
Many of the higher dimensional black holes whose entropy can be computed in 
string theory have near-horizon geometries that look like that of the BTZ 
black hole \cite{Skenderis}, allowing thermodynamic properties to be computed 
by the methods of section \ref{btz}.  If the Virasoro algebra of the BTZ black 
hole at infinity can be related to our near-horizon algebra, it may be possible 
to demonstrate the role of horizon constraints in these string theoretical 
black holes.  A similar interpretation may be possible for induced quantum 
gravity, where a conformal field theory description is also possible \cite{Frolov}.  
Whether a corresponding result exists for loop quantum gravity is an open question.

Third, there is much more to black hole thermodynamics than the Bekenstein-Hawking
entropy.  If the ideas described here are correct, the Goldstone-like ``would-be 
diffeomorphism'' degrees of freedom must couple properly to external fields to 
produce Hawking radiation.  In 2+1 dimensions, there has been one computation 
of this sort \cite{Emparan}, in which the coupling of a classical source to the 
boundary degrees of freedom was shown to yield the correct Hawking radiation.  
A recent discussion of Hawking radiation in terms of diffeomorphism anomalies 
at the horizon \cite{Wilczek} may also be relevant, and it may be possible to
obtain some information by imposing horizon constraints in a Lagrangian
formalism \cite{Mattingly}.  But the investigation of this issue has barely 
begun.

\begin{flushleft}
\bf Acknowledgments
\end{flushleft}
This work was supported in part by the U.S.\ Department of Energy under grant
DE-FG02-99ER40674. This paper is dedicated to Rafael Sorkin on the occasion 
of his sixtieth birthday.  A similar article, based on a related talk, will appear
in the {\it Proceedings of the Fourth Meeting on Constrained Dynamics and Quantum 
Gravity (QG05)}.

\end{document}